\documentclass[
  reprint,
  groupedaddress,  
  amsmath,amssymb,
  aps,
  prb,
]{revtex4-2}

\usepackage{graphicx}
\usepackage{dcolumn}
\usepackage{bm}
\usepackage{hyperref}
\usepackage{enumitem}
\usepackage{soul}

\usepackage[caption=false]{subfig}
\usepackage{multirow}
\usepackage{graphicx}
\usepackage{dcolumn}
\usepackage{bm}
\usepackage{threeparttable}
\usepackage{color}
\usepackage[dvipsnames]{xcolor}
\usepackage{chemformula}
\usepackage{xspace}
\usepackage{orcidlink}
\usepackage{amsmath}
\usepackage{calrsfs}
\usepackage{ulem}
\usepackage{tabularx}
\usepackage{lineno}

\begin{document}

\preprint{APS/123-QED}

\title{Strongly inhomogeneous spin dynamics induced by \\ ultrashort laser pulses with a gradient intensity profile}

\author{T. T. Gareev\,\orcidlink{0000-0002-1678-2444}}
 \email{timur.gareev@ru.nl}
 \homepage{}
\affiliation{Radboud University Nijmegen, Institute for Molecules and Materials, 6525 AJ Nijmegen, The Netherlands}

\author{N. E. Khokhlov\,\orcidlink{0000-0001-7360-1163}}
\affiliation{Radboud University Nijmegen, Institute for Molecules and Materials, 6525 AJ Nijmegen, The Netherlands}

\author{L. Körber\,\orcidlink{0000-0001-8332-9669}}
\affiliation{Radboud University Nijmegen, Institute for Molecules and Materials, 6525 AJ Nijmegen, The Netherlands}

\author{A. V. Kimel\,\orcidlink{0000-0002-0709-042X}}
\affiliation{Radboud University Nijmegen, Institute for Molecules and Materials, 6525 AJ Nijmegen, The Netherlands}

\date{\today}

\begin{abstract}
\textit{}
The optical pump-probe technique is a common tool for investigation of ultrafast spin dynamics, which usually utilizes single-diode detection averaging the dynamics over the pumped area.
Using ultrafast imaging technique, we show experimentally that a femtosecond laser pulse with a gradient distribution of intensity efficiently excites strongly inhomogeneous spin dynamics on spatial scales much smaller than the pump spot size. 
The mechanism responsible for the inhomogeneous distribution is based on temperature gradients and corresponds to a sign change of the torque derivative in different areas of the pump. 
We argue that the observed phenomenon is general for the systems with competitive magnetic anisotropies. 
Overlooking this effect in the majority of pump-probe experiments can result in a dramatic underestimate of the lifetime and amplitude of the laser-induced spin dynamics.
\end{abstract}

\maketitle


The field of ultrafast magnetism aims to understand spin dynamics in magnetically ordered media exited by ultrashort (sub-100 ps) stimuli. 
The ability to generate and control spin waves in magnetic materials using femtosecond laser pulses \cite{van2002all, vomir2005real, kirilyuk2010ultrafast, keatley2011ultrafast, schellekens2014ultrafast,maksymov2015magneto, busse2015scenario, titze2024all, miedaner2024excitation} opened exciting opportunities to merge spintronic or magnonic technologies with photonics \cite{liu2016optomagnonics, dey2021opto,wang2022recent, demirer2022integrated,wang2022picosecond, pezeshki2023integrated}. 
As progress in the field of ultrafast magnetism is predominantly driven by experimental discoveries, the pump-probe technique has practically become the main tool in this field of research. 
Here we report that using ultrafast imaging, which has rarely been used so far in experimental studies of ultrafast magnetism, we discovered a counterintuitive, and therefore often overlooked, strongly inhomogeneous spin dynamics. 
We argue that such a dynamics can be excited even if the laser beam has a relatively large diameter and a Gaussian distribution of light intensity.
In particular, the strongly inhomogeneous magnetization oscillations on spatial scales much smaller than the pump spot originate from the different ultrafast dynamics of competing magnetocrystalline and shape anisotropies. 

\begin{figure}
    \centering
    \includegraphics[width=1\linewidth]{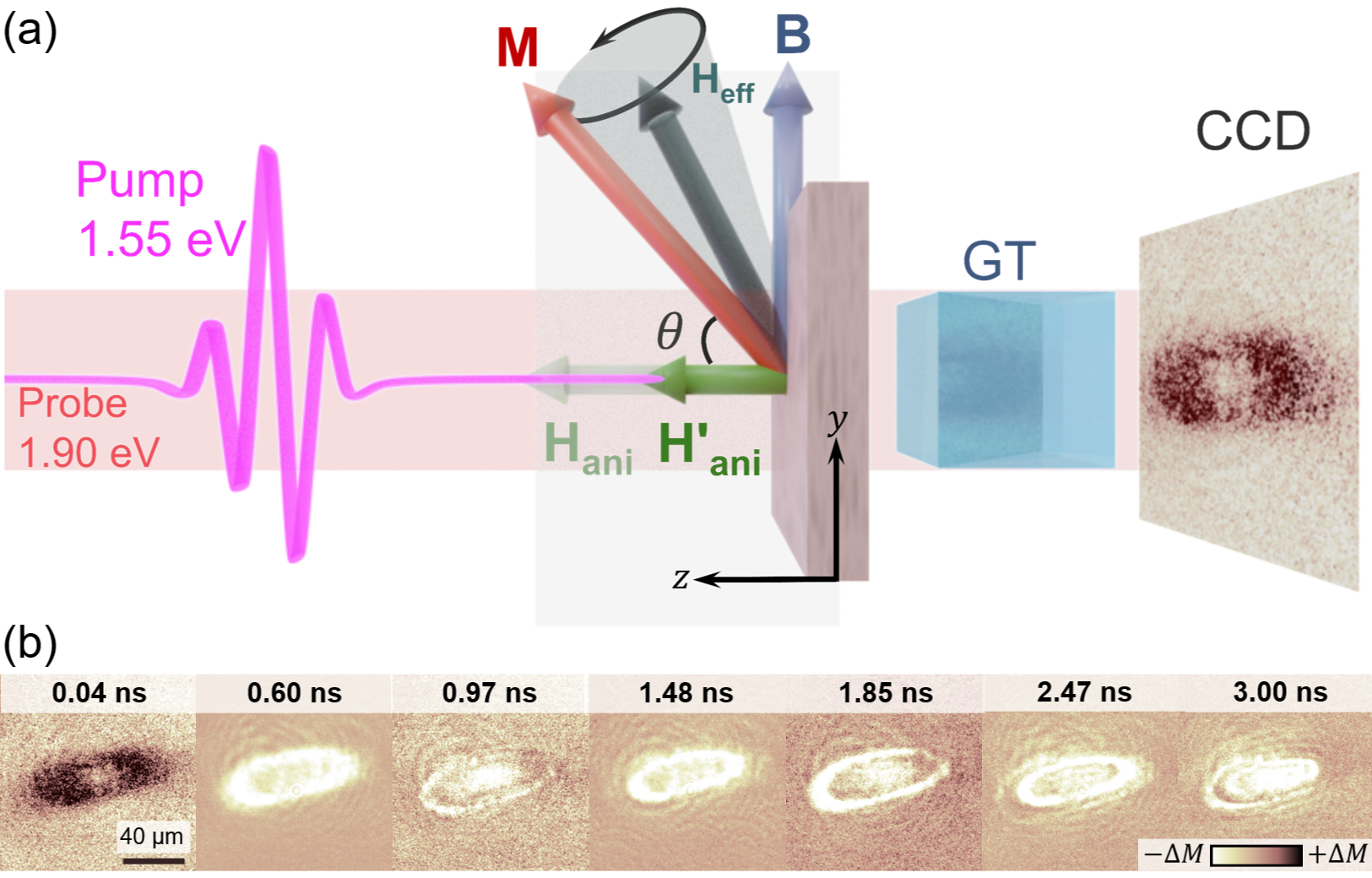}
    \caption{
    (a) Scheme of the experimental geometry. GT - Glan-Taylor polarizer.
    (b) MO snapshots of the spin dynamics at different time moments. 
    The contrast scale was adjusted for each snapshot individually {\color{black}(see details in text)}.
    The images were obtained at magnetic field $B=45$~mT.
    }
    \label{fig:experiment}
\end{figure}

The sample is an epitaxial 1.72\,\(\mu\)m thick film of iron garnet (BiLu)$_3$(FeGa)$_5$O$_{12}$, grown on a Gd$_3$Ga$_5$O$_{12}$ (GGG) substrate with a (110) crystallographic orientation. 
The saturation magnetization of the sample is $13.5$ kA/m and its out-of-plane uniaxial anisotropy has a parameter $194$ J/m.
In the experiments, an external magnetic field \(\mathbf{B}\) is applied in the sample plane using an electromagnet (Fig. \ref{fig:experiment}a). 
We studied magnetization dynamics launched by femtosecond laser pulses using the time-resolved pump-probe imaging technique as described in detail elsewhere \cite{Dolgikh2023}. 
A Ti:Sapphire regenerative amplifier provides linearly polarized 45-fs laser pulses at a photon energy of 1.55\,eV and 1 kHz repetition rate. 
The initial pulse is divided into pump and probe parts. 
The pump was focused on the sample at an incidence angle of \(10^\circ\), forming an elliptical spot. 
The intensity distribution had a Gaussian shape with full width at half maximum (FWHM) of 30 and 130 $\mu$m along minor and major axes, respectively. 
An optical parametric amplifier converted the photon energy of the probe pulses to 1.9\,eV. 
The linearly polarized probe passed through a mechanical delay stage equipped with a retroreflector to enable time resolution.
The probe beam remained unfocused {\color{black} with FWHM of 2 mm and fluence of 0.1 mJ/cm$^2$, yielding negligible, spatially uniform heating (Sec. I in Suppl. Materials).}
A Glan–Taylor polarizer placed between the sample and a CCD camera enabled the detection of the out-of-plane magnetization component \(m_z\) via the magneto-optical (MO) Faraday effect. 
The resulting MO images were recorded {\color{black} as the difference between two images at different pump-probe time delays: one recorded without the pump pulse (serving as background) and one recorded with the pump pulse}.
All experiments were carried out at room temperature.


The geometry with non-collinear orientation of \textbf{B} and anisotropy axis that we used provides the excitation of magnetization dynamics due to ultrafast reduction of magnetocrystalline anisotropy \cite{kirilyuk2010ultrafast, kalashnikova2023ultrafast}.
Here, we measured the excited dynamics in a standard stroboscopic pump-probe technique except that the detector was a camera, not a photodiode or a balanced pair of the two.
As a result, the clear ring pattern is detected at the pump-probe delay $t=0.04$ ns for a pump fluence of 85 mJ/cm$^2$ (Fig. \ref{fig:experiment}b).
{\color{black} Here, the MO contrast scale was individually adjusted for each snapshot and $|-\Delta M| = |+ \Delta M|$.
Thus, the unpumped area is always fixed at $\Delta M_0=0$ and can be used as a reference  (for details see Suppl. Materials, Sec. II).}
In the central area, {\color{black}{the transient change in the MO signal}} is higher, while at the edges it is lower, indicating that the $z$-component of $M_s$ is of opposite signs in the areas.
At 0.6 ns the contrast is the opposite, and at 0.97 ns the contrast is reversed again with a brighter central area. 
This alternating behavior continues for longer times within the pumped area.
Moreover, the number of rings in the pattern increases progressively up to 3 ns (full video of the spin dynamics is available in the Supplemental Materials \cite{SM}).
At the same time, the dynamics stays within the pumped area and does not propagate outside in the form of spin waves {\color{black}{(Suppl. materials, Sec. III})}, as is typically observed for the tightly focused laser pump \cite{satoh2012directional, au2013direct, khokhlov2019optical}. 

\begin{figure}[t!]
    \centering
    \includegraphics[width=0.9\linewidth]{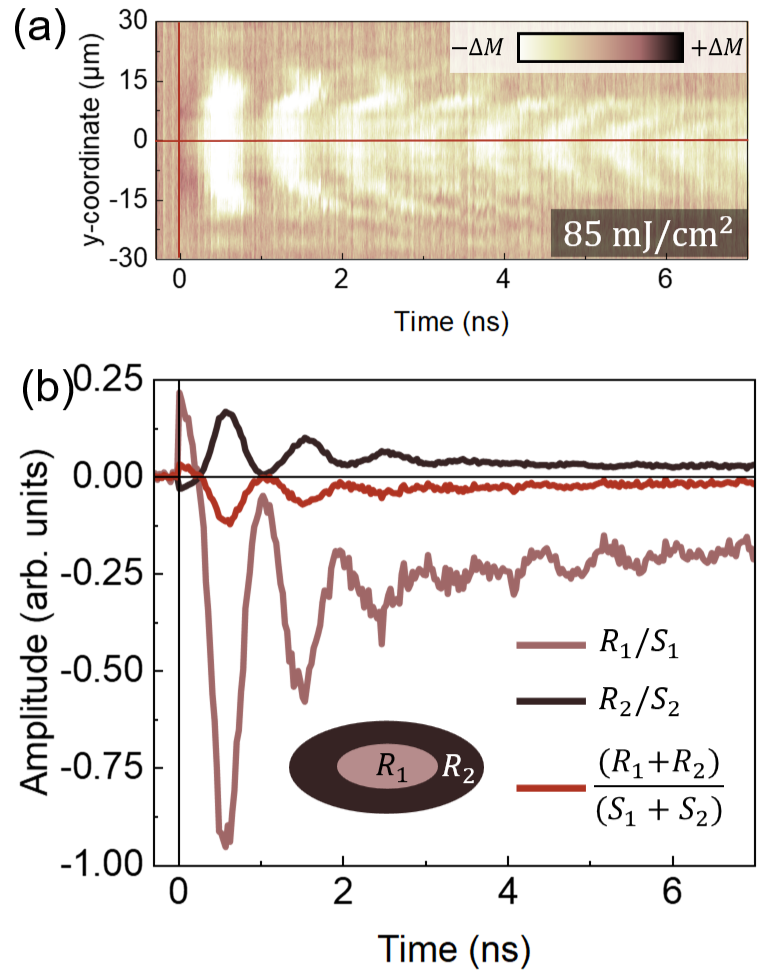}
    \caption{
    (a) Spatial-temporal evolution of the magneto-optical contrast in the central cross-section of the laser-excited area. 
    (b) Magnetization dynamics {\color{black}averaged} over the central area of the spot $R_1/{\color{black}S_1}$ ({\color{black}$S_1$ is an area of the} ellipsis with 17 $\mu$m and 42 $\mu$m axes), over the edge area $R_2/{\color{black}S_2}$ ({\color{black}$S_2$ is an area of the} ellipsis with 33 $\mu$m and 75 $\mu$m axes, exclude $R_1$), and over the total area $(R_1 + R_2)/{\color{black}(S_1+S_2)}$. 
    The laser fluence is 85 mJ/cm$^2$, and $B=45$ mT.
    }
    \label{fig:osc}
\end{figure}

To analyze the spatial-temporal evolution of the spin dynamics, we plotted the cross section {\color{black} along $y$-axis} of the MO snapshots (Fig.~\ref{fig:osc}a). 
The resulting spatial-temporal map demonstrates how the dynamics varies across the pump spot with the gradient of laser fluence. 
It is clearly seen that the phase shift between the central and outer areas increases with time, {\color{black} and the oscillation frequency is higher in the central region of maximum fluence.
It is in agreement with previous studies on laser-induced spin waves with $B$-field orientation along the hard anisotropy axis \cite{khokhlov2019optical, filatov2022spectrum}, when the spatial temperature gradient occurs (Section VIII in Suppl. Materials).
A cross-section along the $x$-axis demonstrates the same positive phase shift (sec. IV in Suppl. Materials).}
To quantify the changes, we extracted the average intensity from the central area of the spot $R_1$ with minor and major axes of 17~$\mu$m and 42~$\mu$m, respectively, and from the surrounding radial area $R_2$ with minor and major axes of 33~$\mu$m and  75~$\mu$m, respectively (excluding $R_1$ region). 
{\color{black} The regions \(R_1\) and \(R_2\) are selected to cover areas where spin dynamics has opposite initial phases at high pump fluence (sec. V in Suppl. Materials).}
As a result, the oscillations in $R_1$ and $R_2$ exhibit opposite initial phases from the very beginning of the dynamics (Fig.~\ref{fig:osc}b). 
Importantly, the rings extend over tens of micrometers in a few picoseconds, suggesting that the dynamics is not driven by any propagation of magnons whose group velocities are typically three orders of magnitude lower \cite{chumak2017magnonic, caretta2020relativistic, mahmoud2020introduction, Jiapeng_AdvElMat2023}.
This indicates that magnetization experiences initial torques acting in mutually opposite directions in different regions of the pumped area because of the inhomogeneous temperature profile of the laser pulse.
{\color{black}Moreover, although spatially averaged data (Fig.~\ref{fig:osc}b) indicate strong damping, single-pixel measurements (Fig.~\ref{fig:osc}a) reveal substantially longer oscillation lifetimes (up to 6.5 ns).
This highlights how spatial averaging, common in standard pump-probe measurements, can obscure intrinsic magnetization lifetimes and amplitudes, underscoring the necessity for spatially resolved detection.}



\begin{figure}[b]
    \centering
    \includegraphics[width=1\linewidth]{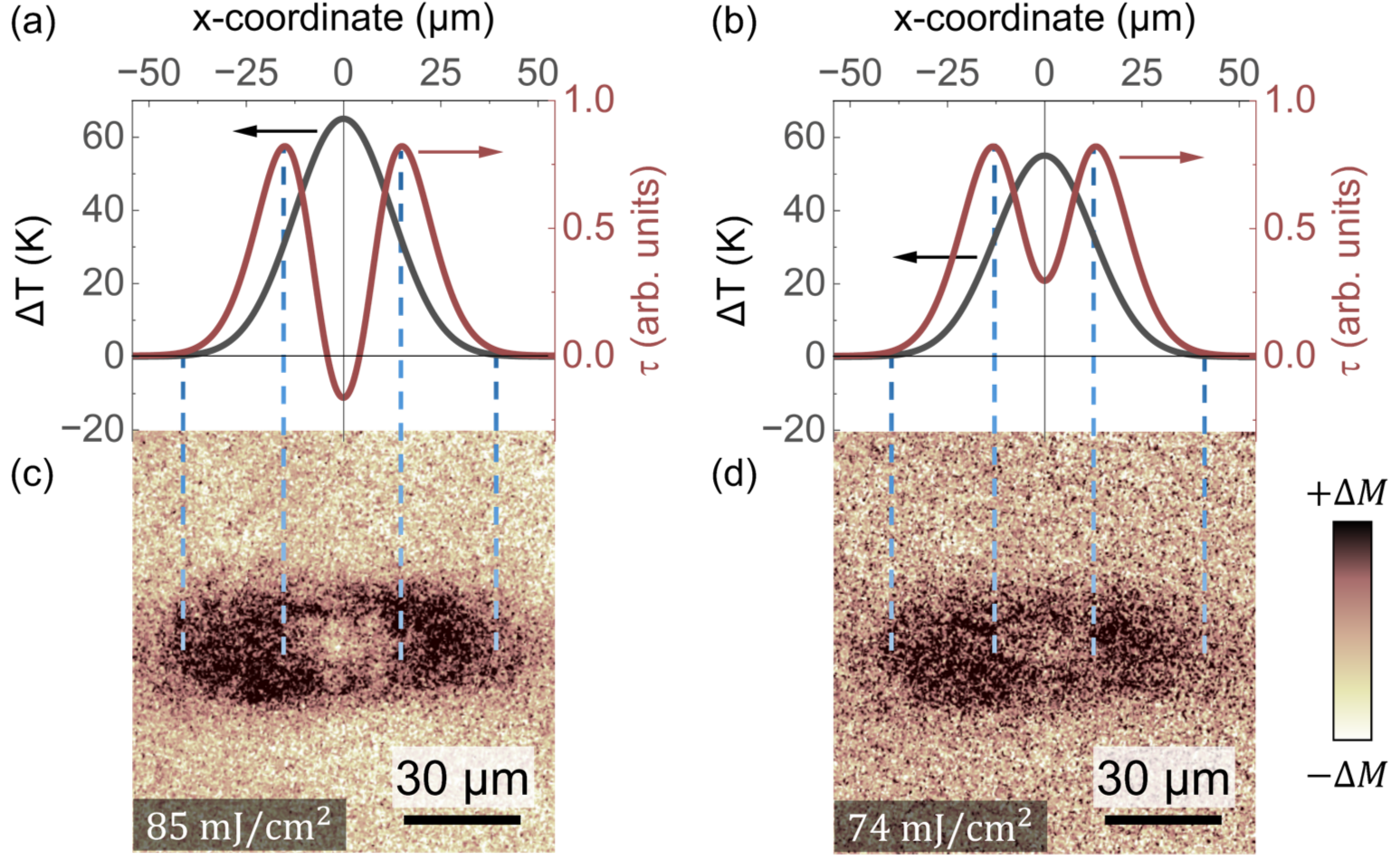}
    \caption{
    (a) Calculated Gaussian pump temperature profile with $\Delta T = 65$~K (black line) and corresponding spatial distribution of the laser-induced torque $\tau$ (red line). 
    (b) Calculated Gaussian pump temperature profile with $\Delta T = 55$~K (black line) and corresponding spatial distribution of $\tau$ (red line). 
    (c) Ring pattern observed in experiment at $t=0.04$~ns with a laser fluence of 85~mJ/cm$^2$ and $B=45$~mT.  
    (d) Ring pattern observed in experiment at $t=0.04$~ns with a laser fluence of 74~mJ/cm$^2$  and $B=45$~mT. 
    }
    \label{fig:td}
\end{figure}

\begin{figure*}[t!]
    \centering
    \includegraphics[width=0.9\linewidth]{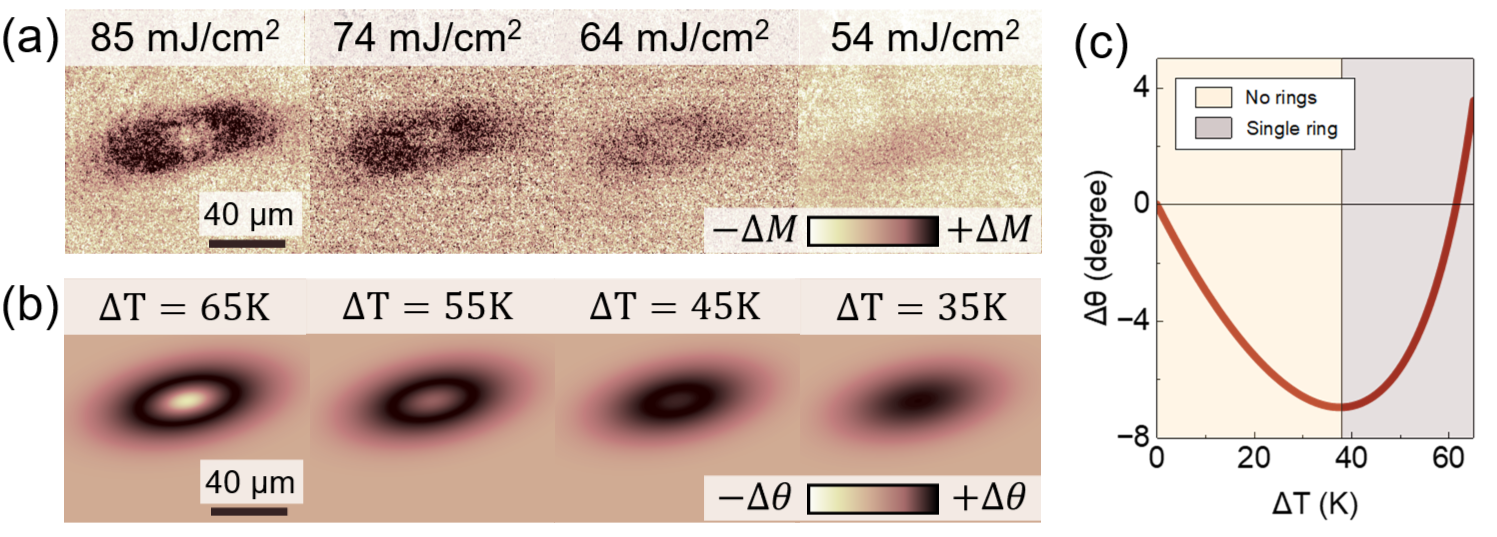}
    \caption{Pump fluence dependence.  
    (a) Experimental snapshots of the spatial distribution of the out-of-plane magnetization component $m_z$ for different pump fluences taken at $t=0.04$~ns.  
    (b) Spatial distributions of the variation of polar angle $\Delta\theta$ calculated as direct minimization of \eqref{eq:TDpotential} for different peak values of pump-induced temperature increase $\Delta T$.
    (c) Calculated variation \(\Delta \theta\) versus \(\Delta T\). 
    Ring formation is observed if derivative ${\partial\theta}/{\partial T}$ changes its sign.
    {\color{black} $\Delta T$ value is estimated similarly as in Ref. \cite{Dolgikh2023} (Suppl. Materials, Sec. IX).}
}
    \label{fig:pwr}
\end{figure*}

To explain the initial torque distribution, we modeled the system as an ensemble of non-interacting macrospins, neglecting any interactions between them such as exchange or dipolar coupling {\color{black} due to fast ring formation, as discussed above}.
In this approximation, we describe each macrospin by a magnetic energy density $\mathcal{F}$ that contains only local uniaxial and effective dipolar anisotropies, as well as the Zeeman interaction:
%
\begin{equation}\label{eq:TDpotential}
    {\mathcal{F}} = \left(K_\mathrm{u} - \frac{1}{2}\mu_0 M_\mathrm{s}^2\right) \sin^2\theta -  M_s B \sin\theta,
\end{equation}
where, $\theta$ is the polar angle of the magnetization vector $\textbf{M}$ (Fig.\ref{fig:experiment}a); \(K_u\) and \(M_s\) are the temperature-dependent uniaxial anisotropy parameter and saturation magnetization, respectively; $B = \mu_0H$ is the magnetic field strength, $\mu_0$ is the vacuum magnetic permeability.
Here, we assume that temperature dependencies $M_s(T)$ and $K_u(T)$ follow an adapted Akulov--Zener approach \cite{akulov1936quantentheorie, zener1954classical}:
\begin{multline}\label{eq:AkulovZener}
    M_s(T) = M_s(0)\left[ 1 - \left( \frac{T}{T_C} \right)^{3/2} \right];\\
    K_u(T) = K_u(0)\left[ 1 - \left( \frac{T}{T_C} \right)^{\beta} \right]^a,
\end{multline}
where $\beta = 12$ following Ref.\,\cite{davies2019anomalously}, and $a=3$ as predicted for single-ion uniaxial anisotropy \cite{zener1954classical, CALLEN_JPCS_1966} and
conserving after laser-induced dynamics in the experiments \cite{Gerevenkov_PhysRevMaterials2021, Blank_AdvMatInter2022}.
{\color{black}We excluded non-thermal ultrafast effects (e.g.\ Cotton–Mouton, modification of magnetic anisotropy, etc.) from the model, as the detected oscillations do not demonstrate a dependence on orientation of pump polarization (Suppl. Materials, Sec. VI).}
Equations \eqref{eq:TDpotential} and \eqref{eq:AkulovZener} give the spatial distribution of the initial torque $\tau = -\partial \mathcal{F}/\partial \theta$ for the Gaussian intensity profile of the pump (Fig.\ref{fig:td}a,b).
{\color{black}This total torque is defined by the interplay of the anisotropies terms and the Zeeman energy with different temperature dependencies \eqref{eq:AkulovZener}.
As a result, the derivative $\partial \tau / \partial x$ changes sign with temperature, and ring formation occurs inside the pumped area (Fig. \ref{fig:td}a,c).}
Therefore, there should be a threshold value for pump-induced temperature changes $\Delta T$ for ring formation.
Indeed, in the experiment, $M_z(r)$ forms the ring if laser fluence exceeds 60 mJ/cm$^2$ (Fig. \ref{fig:pwr}a), and it is well reproducible in simulations (Fig. \ref{fig:pwr}b).
{\color{black}Furthermore, the value of $\tau$ is defined by the deviation of the equilibrium angle $\Delta \theta$ with temperature.}
Thus, the threshold for ring formation is also defined by the change in sign of the derivative $\partial \theta / \partial T$ (Fig. \ref{fig:pwr}c).
We notice that the derivative $\partial \theta / \partial T$  changes sign only once.
This means that the initial MO image will have only one initial ring when the pump fluence and the corresponding $\Delta T$ exceed a threshold value, consistent with the experiments and calculations (Fig.\ref{fig:pwr}).
{\color{black}We notice that the model of the ring formation does not consider ultrafast processes in the first few hundreds of femtoseconds, like demagnetization, electron-phonon thermalization, etc. 
Thus, it is applicable at times larger than a few picoseconds in garnets with the corresponding initial ring formation.}

{\color{black}After the above detailed analysis of the torque and the formation of one initial ring, we would like to address the evolution of other dynamics parameters in the pump-induced gradient of $\Delta T$, as it produces a long-lasting spatial variation of the parameters $K_u$ and $M_s$.
First, changes in the parameters lead to a variation in the ratio of competing out-of-plane uniaxial and in-plane shape anisotropies, as they follow the corresponding temperature dependencies \eqref{eq:AkulovZener}.
As a result, the equilibrium angle $\theta$ at high temperatures (area $R_1$) is close to the orientation in the plane, while $\theta$ in the area $R_2$ tends to be out of the plane (Fig.\ref{fig:pwr}c).}
It leads to the spatial variation of the slow-varying inhomogeneous background of opposite signs in the center $R_1$ and in the outer area $R_2$, observed in the experiments (Fig.\ref{fig:osc}b).

{\color{black}Second, the increase in the number of rings with time is also attributed to the spatial variation of $K_u$ and $M_s$.}
The parameters define the local magnetization precession frequencies, which vary in space, resulting in phase variation across the pump spot at times of a few nanoseconds (Fig.\ref{fig:osc}).
Although the similarity of the ring-like pattern reported here and the final magnetization state after single-pulse magnetization switching ~\cite{davies2019anomalously, peng2023plane, Mishra_PhysRevResearch2023}, our observations reveal a markedly different mechanism at an early stage of the dynamics.
In particular, the ring pattern in Fig.\ref{fig:experiment}b is formed already in 0.04 ns after laser excitation, and we also see it clearly for the integrated area results (Fig. \ref{fig:osc}b), while in Refs. \cite{davies2019anomalously, Mishra_PhysRevResearch2023} laser excitation first led to a homogeneous in sign laser-induced torque, and the rings formed only afterward on a time of precession period. 
Thus, our finding shows the additional torque-based mechanism behind the formation of a ring-like pattern.
Moreover, the magnetization precession reported here and the switching in Refs.~\cite{davies2019anomalously, Mishra_PhysRevResearch2023} are two fundamentally different processes: linear and non-linear.
To compare their amplitudes, we used the MO contrast in Fig.~\ref{fig:experiment}b and the value of the static MO Faraday effect to estimate the amplitude of the local magnetization precession. 
For the case in Fig.~\ref{fig:experiment}b, this amplitude is about 1.5$^\circ$. 
In contrast, the strong anharmonic spin dynamics reported in Ref.~\cite{davies2019anomalously} implied amplitudes well above 10$^\circ$.



In conclusion, we show that spin dynamics excited in a typical pump-probe experiment can be strongly inhomogeneous due to its initial torque distribution, and the inhomogeneity takes place on spatial scales much smaller than the pump spot size.
The torque-based mechanism is very general and could be expected in a wide variety of magnetic materials.
It leads to a situation where probing the whole pumped area gives a decrease in the amplitude of detected magnetization oscillations with lifetimes smaller than in the center of the pump.
Overlooking this effect in pump-probe experiments may result in a dramatic underestimation of the spin precession parameters.
{\color{black}In particular, we recover spatially resolved phase and amplitude information lost in conventional schemes, demonstrating a broadly applicable route to control and correctly interpret ultrafast magnetization dynamics.}

\section*{Acknowledgments}
The authors are grateful to A. P. Pyatakov for the sample provided, L. Shelukhin and Th. Rasing for fruitful discussions, C. Berkhout and K. Saeedi Ilkhchy for technical support.
The work is supported by the European Research Council ERC Grant Agreement No. 101054664 (SPARTACUS) and the research program “Materials for the Quantum Age” (QuMat).
The work of T.T.G. was funded by the European Union’s Horizon 2020 Research and Innovation Program under Marie Skłodowska-Curie Grant Agreement No. 861300 (COMRAD). L.K. gratefully acknowledges funding from the Radboud Excellence Initiative.

\section*{Conflict of Interest Statement}
The authors have no conflicts to disclose.

\section*{Data availability}
The raw data generated in this study and the processed data have been deposited in the Zenodo data base \cite{gareev2025dataset}.

\section*{Author Contributions}
\textbf{Timur T. Gareev}: Investigation (experiment, equal; theory, equal); Data Curation; Visualization; Writing –- original draft (equal); Writing –- review and editing (equal).
\textbf{Nikolai E. Khokhlov}: Investigation (experiment, equal; theory, equal); Methodology (experiment); Writing –- original draft (equal); Writing –- review and editing (equal).
\textbf{Lukas K{\"o}rber}: Investigation (theory); Software, Writing –- review and editing (equal).
\textbf{Alexey V. Kimel}: Conceptualization; Supervision; Writing –- review and editing (equal); Project Administration.


\bibliography{bibliography}

\end{document}